\documentclass[aps,twocolumn, floatfix, prx,longbibliography]{revtex4-1}
\usepackage{graphicx}
\DeclareGraphicsExtensions{.pdf}
\usepackage{amsmath,amssymb,bbold,bm,color}
\usepackage{float,hyperref}
\usepackage{epstopdf}
\usepackage{titlesec}
\hypersetup{colorlinks=true,urlcolor=blue}

\newcommand{\bk}{{\bm k}}

\newcommand{\bsig}{{\bm \sigma}}
\newcommand{\btau}{{\bm \tau}}

\newcommand{\cC}{{\cal C}}

\newcommand{\cT}{{\cal T}}

\newcommand{\cP}{{\cal P}}
\newcommand{\cH}{{\cal H}}

\newcommand{\bee}{\begin{equation}}
\newcommand{\ee}{\end{equation}}

\begin{document}

\title{$d$-mon: transmon with strong anharmonicity }

\author{Hrishikesh Patel, Vedangi Pathak, Oguzhan Can, Andrew C. Potter,
  Marcel Franz}
\affiliation{Department of Physics and Astronomy, and Quantum Matter
  Institute, University of British Columbia, Vancouver, BC, Canada V6T 1Z1}

\begin{abstract} 
We propose a novel qubit architecture based on a planar
$c$-axis Josephson junction between a thin flake $d$-wave
superconductor ($d$SC), such as a  high-$T_c$ cuprate
Bi$_2$Sr$_2$CaCu$_2$O$_{8+x}$, and a conventional
$s$-wave superconductor. When operated in the transmon regime the
device -- that we call ``$d$-mon'' -- becomes insensitive to offset charge
fluctuations and, importantly, exhibits at the same time energy level spectrum with strong
anharmonicity that is widely tunable 
through the device geometry and applied magnetic flux. Crucially, unlike previous
qubit designs based on $d$-wave superconductors  the proposed device operates
in a regime where quasiparticles are fully gapped and can be
therefore expected to achieve long coherence times.

\end{abstract}

\date{\today}
\maketitle

{\em Introduction --} Transmon qubit, based on a
superconducting Josephson junction shunted by a large capacitance \cite{Koch2007}, is the workhorse component powering the majority of intermediate scale quantum computers
currently in operation \cite{Martinis2019,IBM2023,Rigetti2023}.
Its chief advantage over other superconducting
qubit architectures is the insensitivity of its active energy levels to the
fluctuations in the offset charge $n_g$ that are typically difficult or
impossible to control. This insensitivity, however, comes at a price: the energy
spectrum of the transmon is only weakly anharmonic which imposes limits
on the speed of operation due to the possibility of the escape from
the code space formed by the two lowest energy eigenstates \cite{SCQubitReview2019,Krantz2020,SiddiqiReview2021}.

We propose here a transmon variant that retains the offset charge
insensitivity of the original device but has an arbitrarily large and
easily tunable energy level anharmonicity. The key to this advance is usage
of a Josephson junction between superconductors with orhogonal order
parameter symmetries. In this paper we consider specifically junctions
comprised of a  $d$-wave and an $s$-wave superconductor, but the idea is applicable
more generally. As is well known ordinary Cooper pair tunneling is
symmetry-prohibited across a $c$-axis $d/s$ Josephson  junction \cite{Klemm2000}. The leading
process that enables passage of supercurrent is co-tunneling of two
Cooper pairs that results in the anomalous $\pi$-periodic current-phase
relation (CPR), $I(\varphi)\simeq I_{c2}\sin{(2\varphi)}$. We will
demonstrate below that the underlying
$\pi$-periodic Josephson free energy $F(\varphi)$ and its two
degenerate minima at $\varphi=\pm\pi/2$ enable the above mentioned key
features of $d$-mon qubit.

\begin{figure}[t]
\includegraphics[width = 8.0cm]{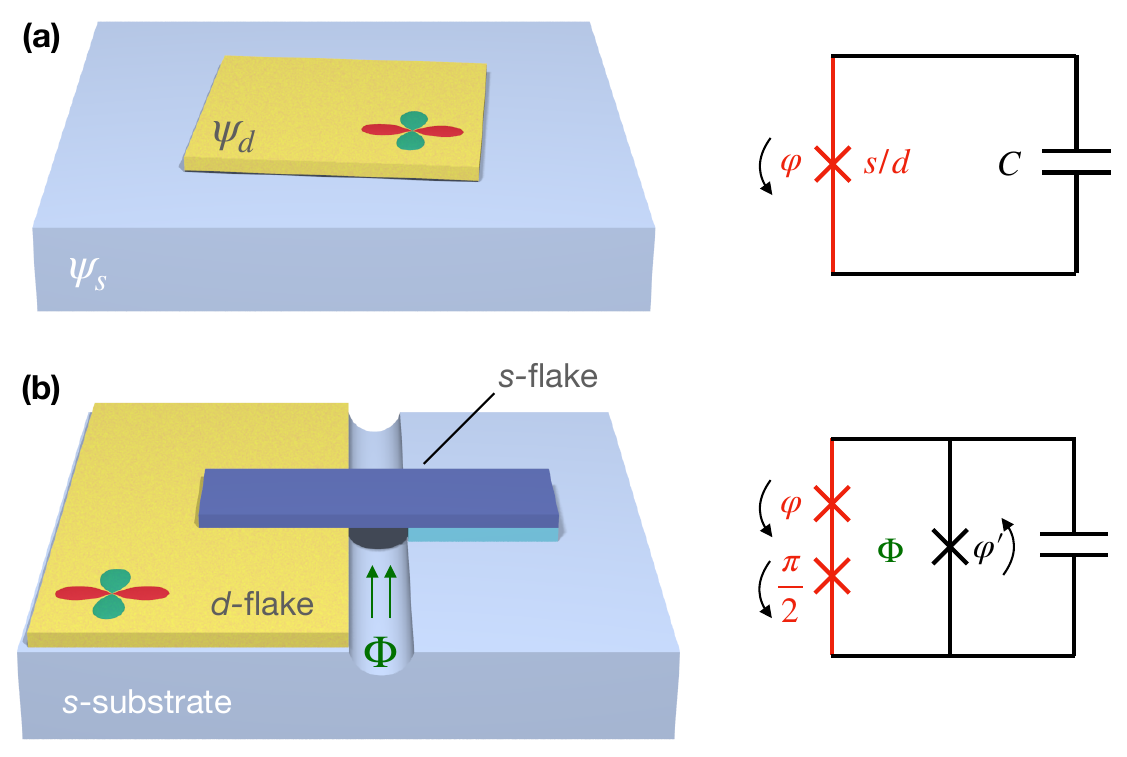}
\caption{Schematic of the proposed $d$-mon device. a) Basic $d$-mon
  architecture and its circuit representation with one $s/d$ junction
  and a capacitor $C$.
  b) Split $d$-mon: A large $d$SC flake resting on top of an
  $s$-wave substrate. A small $s$-wave flake bridges the gap threaded
  by magnetic flux $\Phi$. The equivalent circuit contains two $s/d$
  junctions, one ordinary $s/s$ junction and a capacitance $C$.  
}\label{fig1}
\end{figure}
The basic $d$-mon design is illustrated in Fig.\ \ref{fig1}(a) and consists of a
very thin (several monolayers) $d$-wave flake resting on a large
$s$-wave superconducting substrate. High-$T_c$ cuprate
Bi$_2$Sr$_2$CaCu$_2$O$_{8+x}$ (BSCCO) is a well established $d$SC
which has been recently exfoliated down to a monolayer
thickness (while retaining its high critical temperature $\sim
90$K) \cite{Yuanbo2019} and would be a natural material for the flake. As a matter of
principle the $s$-wave substrate can be fabricated of any conventional
superconductor. However, as we discuss in more detail below, a
material compatible with BSCCO -- in that it can proximity-induce a
significant nodal gap -- is required for practical qubit operation.

The Ginzburg-Landau (GL) free energy of the system depicted in Fig.\ \ref{fig1}(a)
can be written as
\begin{eqnarray}\label{f1}
  F[\psi_s,\psi_d] &=&F_s[\psi_s]+F_d[\psi_d] +A|\psi_s|^2|\psi_d|^2 \\
  &+&B(\psi_s\psi_d^\ast +{\rm c.c.}) +D(\psi_s^2{\psi_d^\ast}^2+ {\rm
      c.c.}), \nonumber
\end{eqnarray}
where $\psi_{s/d}$ are complex scalar order parameters and $F_{s/d}$
denote GL free energies of the individual superconductors.
If both superconductors obey tetragonal symmetry then,
importantly, the coefficient $B$ is required to vanish. This is
because under $C_4$ rotation $\psi_s\to \psi_s$ while $\psi_d\to
-\psi_d$. In this situation the leading Josephson coupling arises from
the last term in Eq.\ \eqref{f1}, which is allowed by symmetry and
represents coherent tunneling
of two Cooper pairs across the junction. Denoting the phase
difference between two order parameters by $\varphi$ the resulting
Josephson free energy becomes
\begin{equation}\label{f2}
  F(\varphi)=F_0+2D|\psi_s|^2|\psi_d|^2\cos{2\varphi},
\end{equation}
where $F_0$ contains terms independent of $\varphi$. We note that
although symmetry alone does not fix the sign of $D$ many
microscopic models, including the standard weak-coupling BCS theory,
give $D>0$ which leads to the free energy landscape with two minima at
$\varphi=\pm\pi/2$. In the classical equilibrium the system will spontaneously break the time-reversal symmetry
$\cT$ and chose
one of the minima. In this $\cT$-broken phase the flake can be thought of as a $d\pm is$ 
superconductor.

In reality BSCCO and other high-$T_c$ cuprates, such as
YBa$_2$Cu$_3$O$_{7-x}$ (YBCO), are weakly
orthorhombic (that is, $C_4$ is weakly broken down to $C_2$). In this
case coefficient $B$ in the free energy \eqref{f1} will be non-zero
but we expect it to be small such that the Josephson free energy is
still dominated by the $\cos{2\varphi}$ term. The $B$-term gives
conventional $2\pi$ periodic contribution to $F(\varphi)$ proportional
to $\cos{\varphi}$. At the level of the Josephson free energy the physics
of $d$-mon is therefore similar to the twisted bilayer of $d$-wave
flakes which has been predicted to form a $\cT$-broken $d\pm id'$
phase at twist angles close to 45$^\circ$ \cite{Can2021}. Experimental evidence
for such a state has been recently reported in twisted BSCCO
bilayers \cite{Zhao2021}. Some important differences between the two setups include
the fact the relative strength of the $\cos{\varphi}$ in twisted $d$-wave
bilayers can be controled by the twist angle $\theta$. Also, the $d\pm id'$
phase is topological (characterized by Chern number $\pm 2$), and has gapless topological edge modes that may act as a source of decoherence for a qubit device. By contrast, the
$d\pm is$ phase that is adiabatically
connected to a pure $s$-wave superconductor, is topologically trivial.

{\em Anharmonicity from double-well free energy --}  Taking into
account the junction charging energy the Hamiltonian for the $d$-mon
can be written as
\begin{equation}\label{h1}
 H_\varphi=4E_C(\hat{n}-n_g)^2+E_J\cos{2\varphi} +\delta U(\varphi),
\end{equation}
where $E_C=e^2/2C$ is the charging energy of the junction with
capacitance $C$, $\hat{n}=-i\partial_\varphi$ is the Cooper pair
number operator, $E_J$ denotes the junction Josephson energy, and $n_g$ is the offset charge. The
potential $\delta U$ is given by 
\begin{equation}\label{h2}
  \delta U(\varphi) = -\eta E_J
    (\cos{\phi_{\rm ex}}\cos{\varphi}-\sin{\phi_{\rm ex}}\sin{\varphi}).
\end{equation}
The first term in $\delta U$ represents the residual single-pair tunnelling caused e.g.\
by weakly broken $C_4$ symmetry discussed above. The second term breaks $\cT$
explicitly -- it makes the double-well asymmetric -- and could arise
from external magnetic flux $\Phi$ in the split $d$-mon design depicted
in Fig.\ \ref{fig1}(b) and discussed in more detail below. We are interested in the transmon regime characterized by
$E_J\gg E_C$ and a situation when $\delta U$ represents a small perturbation,
that is, $|\eta|\lesssim 1$.

\begin{figure}[t]
\includegraphics[width = 8.6cm]{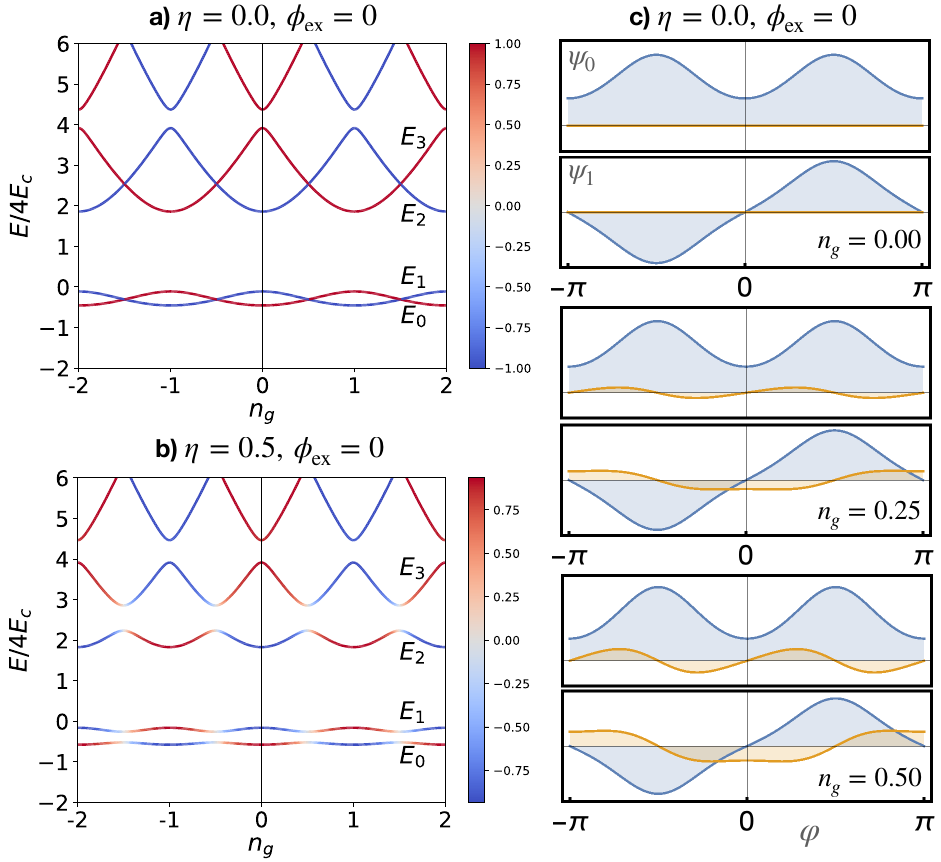}
\caption{Energy levels and wavefunctions of the $d$-mon Hamiltonian
  \eqref{h1} for $E_J/E_C=8$. (a,b) Lowest energy levels as a function of offset charge $n_g$. The color represents parity $\langle\cP\rangle \in (-1,1)$. (c) Wavefunctions
  $\psi_0(\varphi)$ and  $\psi_1(\varphi)$ 
  belonging to the two lowest energies for selected values of $n_g$. Blue
  (orange) lines represent real (imaginary) parts of $\psi_{j}$.
}\label{fig2}
\end{figure}
Consider first the case $\eta=0$. In this limit it is easy
to see that the Hamiltonian \eqref{h1} conserves the Cooper parity
$\cP=(-1)^{\hat{n}}$. In each parity sector the eigenstates and
energy eigenvalues can be obtained analytically in terms of Mathieu
functions as originally discussed in Ref.\ \cite{Koch2007}. In
the transmon regime $E_J\gg E_C$ an
accurate approximation for the splitting between two lowest energy levels can be derived
\cite{Smith2020},
\begin{equation}\label{h3}
  \Delta E\simeq 16 E_C\sqrt{2\over \pi}\left({2E_J\over E_C}\right)^{3/4}
  e^{-\sqrt{2E_J/E_C}}\cos{(\pi n_g)}.
 \end{equation} 
 Band crossings at half-integer values of $n_g$ in Eq.\ \eqref{h3} are
 exact and follow from parity conservation.
\begin{figure*}[t]
\includegraphics[width = 17.6cm]{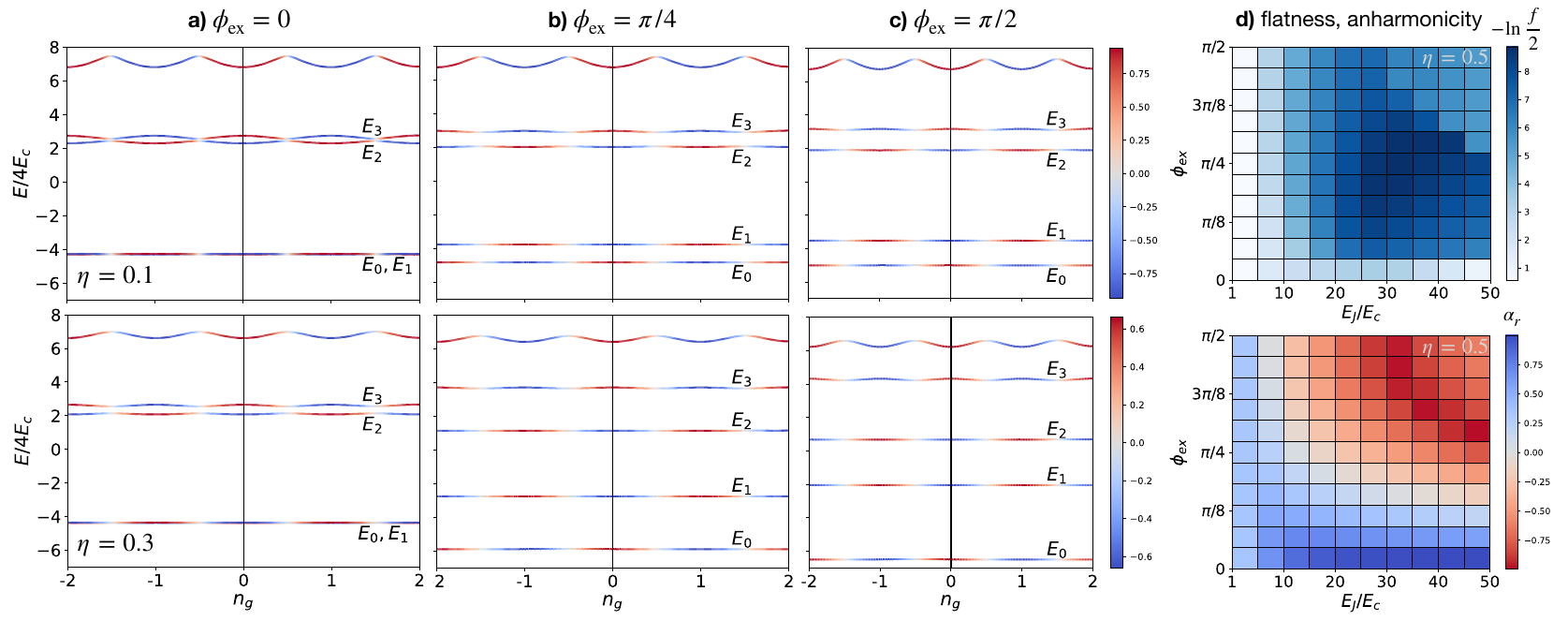}
\caption{Characteristic $d$-mon properties. (a-c) Lowest energy eigenvalues as a function of the offset charge $n_g$ for $E_J/E_C=32$ and representative values of parameters $\eta$ and $\varphi_{\rm ex}$. d) Band flatness $f$ and relative level anharmonicity $\alpha_r$, both defined in the text. Large values of $-\ln{(f/2)}$ indicate very flat bands whereas values of $\alpha_r$ away from 0 signal anharmonicity, which can be both positive or negative. 
}\label{fig3}
\end{figure*}

 We will be interested
 primarily in the case of nonzero $\eta$ when $\cP$ conservation no
 longer applies. In this case analytical results are not available
 but it is possible to solve the problem numerically by representing the
 Hamiltonian \eqref{h1} as a matrix in the Cooper pair number basis
 $|n\rangle =e^{i n\varphi}/\sqrt{2\pi}$ with $n$ 
 integer. We truncate the infinite Hamiltonian matrix $H_{nm}=\langle
 n|H|m\rangle$ according to $|n|,|m|\leq n_{\rm
   max}$ and diagonalize the resulting matrix of size $2 n_{\rm
   max}+1$.  As shown in Fig.\ \ref{fig2}(a) for
 $\eta=0$ the energy levels behave in accord with the analytical
 result Eq.\ \eqref{h3}. Note that the relevant wavefunctions
 $\psi_j(\varphi)$ with $j=0,1$ can be classified as symmetric and antisymmetric
 with respect to the $\varphi=0$ origin only for integral $n_g$. More
 generally they also contain an imaginary part that has opposite
 symmetry, Fig.\ \ref{fig2}(c). 

When $\eta\neq 0$ the parity-protected energy crossings are lifted, Fig.\ \ref{fig2}(b). As a result, the low-lying energy bands flatten out -- the qubit becomes insensitive to the offset charge fluctuations in the same
 way as the original transmon. Importantly, in $d$-mon this
 feature does {\em not} come at the expense of anharmonicity. This is illustrated in Fig.\ \ref{fig3}: Although the
 detailed behavior depends on the relative amplitude of the two terms comprising $\delta U$, in nearly all cases, one can achieve large
 anharmonicity, while the energy levels remain essentially independent of $n_g$, as measured by the flatness indicator $f$ defined as $f=w_1/\omega_{10}$, where $w_j$ denotes the bandwidth of the $j$-th band and $\omega_{ij}=E_i-E_j$. The anharmonicity is measured by parameter $\alpha_r=(\omega_{21}-\omega_{10})/\omega_{20}$. Inspection of Fig.\ \ref{fig3}(d) reveals that for $E_J/E_C\gtrsim 20$ bands become extremely flat and by adjusting the flux $\phi_{\rm ex}$ one can always achieve significant level anharmonicity. The limit of zero flux is special: here the bands become flat but remain nearly degenerate. This is because, as discussed in SM, $\langle\psi_1|\cos{\varphi}|\psi_0\rangle=0$ and hence $\delta U$ has no effect to leading order in perturbation theory. On the other hand $\langle\psi_1|\sin{\varphi}|\psi_0\rangle\neq 0$ which implies that $\delta U$ is much more effective at splitting the bands at non-zero $\phi_{\rm ex}$. Additional results characterizing various parameter regimes of the Hamiltonian \eqref{h1} are given in SM.

 {\em Quasiparticles --} Because of their reliance on $ab$-plane junctions a significant drawback of some earlier cuprate-based qubit designs  \cite{Zagoskin1999,Ioffe1999,Blais2000,Amin2005}
 was the presence of
 quasiparticles that survive in the vicinity of the Dirac nodes in their
 $d$-wave order parameter down to arbitrarily low energies
 \cite{Fominov2003,Amin2004}. In the
 present $d$-mon architecture Dirac nodes are gapped out due to the
 proximity effect. As we now explain the thin $d$SC flake in Fig.\ \ref{fig1}(a)
 becomes a $d\pm is$ superconductor whose quasiparticles are gapped
 everywhere on its Fermi surface.

 The microscopic Hamiltonian for  electrons near the $d/s$
 interface can be written as $\cH_{\rm el}=\sum_\bk\Psi_\bk^\dag H_\bk\Psi_\bk$
 with $\Psi_\bk=(c_{\bk\uparrow},c^\dag _{-\bk\downarrow};
 s_{\bk\uparrow},s^\dag _{-\bk\downarrow} )^T$ and
\begin{equation}\label{h4}
  H_\bk=
  \begin{pmatrix}
    \xi_\bk & \Delta_\bk & t_\bk & 0 \\
    \Delta_\bk & -\xi_\bk & 0 & -t_\bk \\
    t_\bk & 0 & \xi^s_\bk & e^{i\varphi}\Delta_s \\
    0 & -t_\bk  & e^{-i\varphi}\Delta_s & -\xi^s_\bk
 \end{pmatrix}.
 \end{equation} 
 Here $c^\dag _{\bk\sigma}$, $s^\dag_ {\bk\sigma}$ denote electron
 creation operators in $d$ and $s$ layers, respectively, $\xi_\bk$,
 $\xi^s_\bk$ are the corresponding normal-state dispersions referenced
 to the Fermi level $\mu$ and $\Delta_\bk=\Delta_d\cos(2\alpha_\bk)$ is
the $d$-wave gap function with $\alpha_\bk$ the polar angle
of the momentum vector $\bk$; $\Delta_s$ denotes the $\bk$-independent $s$-wave gap.

We now imagine integrating out the gapped fermion degrees of freedom
in the $s$-wave layer, assuming weak interlayer coupling $t_\bk$  (see
SM for the details of the procedure). The resulting effective
Hamiltonian for the remaining $c$ fermions takes the form
 \begin{equation}\label{h5}
\cH_{\rm eff}=\sum_\bk\psi^\dag_\bk
  \begin{pmatrix}
    \tilde\xi_\bk & \Delta_\bk + e^{i\varphi}m_s\\
    \Delta_\bk+ e^{-i\varphi}m_s& -\tilde\xi_\bk
 \end{pmatrix} \psi_\bk,
 \end{equation} 
 with $\psi_\bk=(c_{\bk\uparrow},c^\dag _{-\bk\downarrow})^T$. The tilde
 on $\xi_\bk$ means that the bare dispersion has been modified
 while $m_s\approx t^2/\Delta_s$ denotes the proximity induced gap.  For a
 fixed classical phase $\varphi$ the quasiparticle spectrum of
 $\cH_{\rm eff}$ reads
 \begin{equation}\label{h6}
E_\bk=\pm\sqrt{
    \tilde\xi_\bk^2+ |\Delta_\bk + e^{i\varphi}m_s|^2}.
 \end{equation} 
Classically, in the absence of fluctuations, the system will reside in
one of the minima of the free energy \eqref{f2} with
$\varphi=\pm\pi/2$. This results in the $d\pm is$ superconductor with a
minimum gap $m_s$ to all quasiparticle excitations.

We see that at the mean-field level (that is, neglecting fluctuations
in the phase $\varphi$) $d$-mon is protected from
quasiparticle poisoning by the proximity gap $m_s$. Of
course in order for the device to function as a useful qubit we must
allow for $\varphi$ to undergo quantum fluctuations. Mathematically,
we need to reintroduce the charging energy $E_C$ and permit the phase
variable to tunnel between the two potential minima. 

An important
question thus arises: Will the quasiparticle gap survive in the
presence of such quantum fluctuations? We tackle this question in SM.
Employing two different methods (a perturbative treatment and a more elaborate semiclassical dilute instanton
approximation) we conclude that the
quasiparticle gap survives even though it is suppressed by
the phase fluctuations according to $m_s\to \tilde{m}_s$ with
 \begin{equation}\label{h7}
\tilde{m}_s=m_s\left|\langle
\psi_0|\sin{\varphi}| \psi_1\rangle\right|
\simeq m_s e^{-\sqrt{E_C/4E_J}},
 \end{equation} 
 where $\psi_j(\varphi)$ are eigenstates of the phase Hamiltonian
 \eqref{h1}. Evaluation of the matrix element shows that phase
 fluctuations lead to a relatively modest gap suppression (between
 5-20\%) in the transmon regime. Intuitively, this can be understood by
 considering the structure of wavefunctions $\psi_j(\varphi)$ depicted in
 Fig.\ \ref{fig2}: Thinking semiclassically the phase particle spends most of its time in the
 vicinity of the classical minima at $\pm\pi/2$ where the
 quasiparticle gap is maximal and only makes short excursions to the
neighborhood of $\varphi=0,\pi$ where the quasiparticles are gapless. The $e^{-\sqrt{E_C/4E_J}}$ factor reflects small quantum fluctuations about these minima. 
Larger values of $E_J/E_C$ (likely relevant for practical implementations of the $d$-mon) suppress these fluctuations and concentrate the phase wavefunctions near the minima, leaving negligible amplitude near $\varphi=0,\pi$.

 {\em Split d-mon --} In the basic $d$-mon realization depicted in Fig.\
 \ref{fig1}(a) it is possible to control some system parameters by adjusting
 the flake size. This follows from the fact that $E_J\sim \mathcal{A}$, the
 interface area, $\mathcal{A}$, while $E_C\sim \mathcal{A}^{-1}$. On the other hand, in the absence of explicit $\cT$ breaking $\phi_{\rm ex}$ is fixed to zero and the $\eta$
 parameter is set by the material properties of the junction and cannot be easily controlled. Hence this basic design can access only a small portion of the parameter space afforded by Hamiltonian \eqref{h1}.

 To gain more flexibility we take inspiration from the split transmon
 architecture \cite{Koch2007} and Ref.\ \cite{Ioffe1999}
 and consider a 3-junction device depicted
 in Fig.\ \ref{fig1}(b). We assume that the large $s/d$ junction has
 negligible charging energy and its phase is therefore permanently locked to one
 of the $\cT$ breaking minima (hereafter we assume $+\pi/2$ for
 concreteness). The charging energy of the small $s$-wave flake is
 non-negligible and phases $\varphi$ and $\varphi'$ are allowed to
 fluctuate. In the small inductance limit the loop cannot trap any
 self-induced flux and the two phases are constrained by the
 single-valuedness of the wavefunction such that
 $\varphi+\varphi'+\pi/2=2\pi\Phi/\Phi_0$, where $\Phi_0=hc/2e$ denotes
 the flux quantum. If we define $\phi_{\rm ex}=2\pi\Phi/\Phi_0-\pi/2$ then
 the total Josephson energy can be written as
 $U(\varphi)=E_J\cos{2\varphi}-E_S\cos(\phi_{\rm ex}-\varphi)$ where
 $E_S$ denotes the Josephson energy of the $s/s$ junction. By
 expanding the second cosine one finds that $U(\varphi)$ coincides with
 the potential in the Hamiltonian \eqref{h1} if one identifies
 $\eta=E_S/E_J$. We conclude that the split $d$-mon architecture Fig.\
 \ref{fig1}(b) is capable of realizing the entire range of properties represented by  Hamiltonian Eq.\ \eqref{h1}.   

{\em Conclusions --} A Josephson junction formed at the interface between $d$- and $s$-wave superconductors displays anomalous $\pi$-periodic CPR which underlies its proposed functionality as an improved transmon qubit. We demonstrated that the resulting ``$d$-mon" is robust to offset charge fluctuations while at the same time exhibits a large and easily tunable level anharmonicity. Importantly, unlike many previous proposals based on $d$SCs, we showed that $d$-mon operates in the regime of fully gapped quasiparticles which is essential to prevent decoherence effects.  While these conclusions are based on a high-level symmetry analysis and modelling it is clear that several conditions on system parameters must be met for the device to function as a practical qubit. At a minimum, one requires a high-transparency $d/s$ interface that would facilitate a significant Cooper pair co-tunneling amplitude and proximity-generate a sizeable nodal gap $m_s$ to protect against quasiparticle poisoning. 

Even though $c$-axis junctions formed of various high-$T_c$ cuprates have been extensively studied \cite{Eckstein1995} -- culminating in recent works on twisted BSCCO junctions \cite{Zhao2021,Lee2021,Zhu2021,Yejin2023,Martini2023} -- not much recent effort went into  experimental studies of $c$-axis $d/s$ junctions. Early experiments on $c$-axis junctions between cuprates (YBCO or BSCCO) and Pb   \cite{Dynes1994,Clarke1996,Kleiner1999} showed only conventional $2\pi$ periodic CPR and were interpreted as evidence for $s$-wave superconductivity in cuprates. Given the obvious contradiction with the present-day consensus on $d$-wave symmetry \cite{Kirtley2000}, and the technological potential of $\pi$-periodic junctions discussed here, it is clearly important to revisit these results using modern techniques designed for ultraclean junction preparation \cite{Zhao2021,Yejin2023,Martini2023}.    An interesting possibility would be to explore interfaces between high-$T_c$ cuprates and iron-based superconductors such as LaFeAsO$_{1–x}$F$_x$ or SmFeAsO$_{1–x}$H$_x$. The latter are layered tetragonal materials with $s$-wave superconducting gap and critical temperatures of up to 50K \cite{Hosono2018}. In addition they exhibit lattice constants ($\sim 4$\AA) similar to cuprates thus potentially enabling fabrication of atomically clean $c$-axis junctions by means of mechanical exfoliation \cite{Yuanbo2019} or atomic layer-by-layer molecular beam epitaxy \cite{Bozovic2021}. 

We note in closing that the $\pi$-periodic CPR that lies at the heart of our $d$-mon proposal can also be achieved in $d$SC bilayers with a near-$45^\circ$ twist angle \cite{Vool2023} or, using more conventional ingredients \cite{Douçot_2012,Bell2014}, as recently implemented in voltage-controlled semiconductor nanowire Josephson junctions \cite{Marcus2020}. These platforms could also be harnessed to produce an improved transmon with large anharmonicity.

{\em Acknowledgments --} The authors are indebted to C.-K.\ Chiu, P.\ Kim, G.\ Refael,  N.\ Poccia, U.\ Vool  
for stimulating discussions and correspondence. The work
was supported by NSERC, CIFAR and the Canada First Research Excellence Fund, Quantum Materials and Future Technologies Program. 
A.C.P. was supported by the US Department of Energy DOE DE-SC0022102, and in part by
the Alfred P. Sloan Foundation through a Sloan Research
Fellowship. M.F.\ and A.C.P. thank Aspen Center for Physics where part of this work was completed.

\bibliography{d-mon-lit}

\newpage
~
\newpage

\newcommand{\<}{\langle}
\newcommand{\e}{\varepsilon}
\newcommand{\up}{\uparrow}
\newcommand{\down}{\downarrow}
\newcommand{\Up}{\Uparrow}
\newcommand{\Down}{\Downarrow}
\renewcommand{\>}{\rangle}
\renewcommand{\(}{\left(}
\renewcommand{\)}{\right)}
\renewcommand{\[}{\left[}
\renewcommand{\]}{\right]}
\renewcommand{\v}[1]{\boldsymbol{#1}} 
\newcommand{\dslash}{d \hspace{-0.8ex}\rule[1.2ex]{0.8ex}{.1ex}}
\renewcommand{\d}{\partial}
\newcommand{\del}{\nabla}
\renewcommand{\div}{\nabla\cdot}
\newcommand{\curl}{\nabla\times}
\newcommand{\eps}{\epsilon}
\newcommand{\p}{\parallel}
\newcommand{\U}{\mathcal{U}}

\appendix

\setcounter{figure}{0}
\setcounter{table}{0}
\setcounter{page}{1}
\makeatletter
\renewcommand{\thefigure}{S\arabic{figure}}

\onecolumngrid
\vspace{0.5cm}
\vspace{0.5cm}
\begin{center}
\bf \large Supplementary Material for ``Transmon with strong anharmonicity''
\end{center}
\vspace{0.5cm}
\twocolumngrid

\section{Additional results for $d$-mon phase Hamiltonian}
\begin{figure*}[t]
\includegraphics[width = 14.6cm]{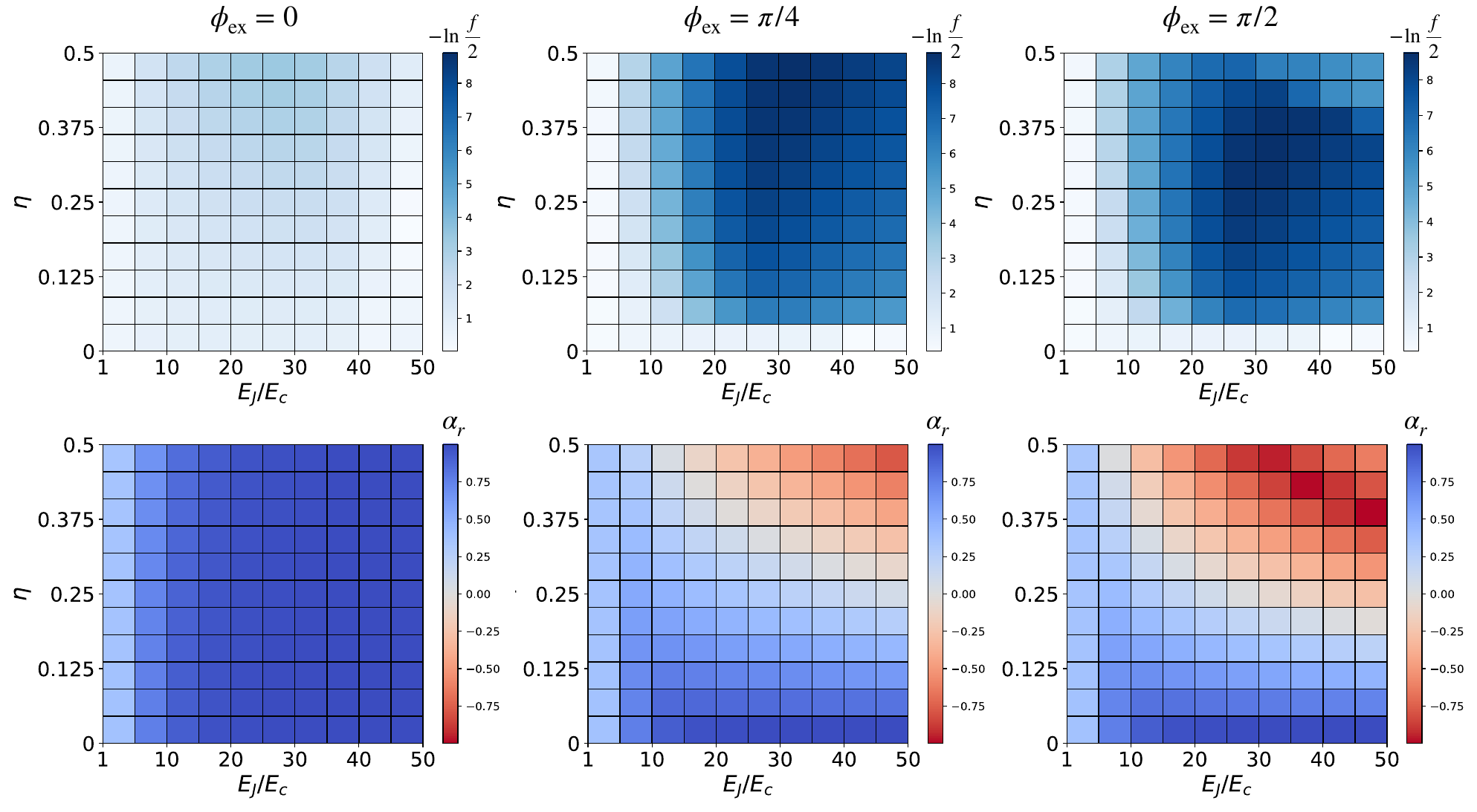}
\caption{Additional numerical results for the band flatness $f$ and the relative anharmonicity $\alpha_r$, both defined in the main text below Fig.\ (3). Top row shows $-\ln(f/2)$ as a function of parameters $\eta$ and $E_J/E_C$ for three values of $\phi_{\rm ex}$ indicated. Bottom row displays $\alpha_r$ in a similar way.}\label{figS0}
\end{figure*}
Fig.\ \ref{figS0} offers a coplementary view of our results for the $d$-mon phase Hamiltonian Eq.\ (3). Once again we observe that a qubit with energies  insensitive to offset charge fluctuations and large anharmonicity, as measured by the band flatness parameter $f$ and the relative anharmonicity $\alpha_r$, respectively, can be achieved for many parameter combinations. The best results, {\em i.e.} large $-\ln(f/2)$ combined with $\alpha_r$ significantly away from zero, are obtained when $\phi_{\rm ex}\neq 0$, a situation that naturally occurs in the split $d$-mon architecture Fig.\ 1(b). It is to be noted that due to the embedded $\pi/2$ junction the split $d$-mon realizes the $\phi_{\rm ex}=\pi/2$ case already at zero external flux. We also remark that in the split $d$-mon, parameter $\eta=E_S/E_J$ can be controlled by adjusting the area of the $s/s$ junction relative to the small $d/s$ junction. Hence, these results indicate that by tuning the external flux and $\eta$ essentially {\em any} desired combination of the band flatness and anharmonicity, either positive or negative, can be achieved.

\section{Derivation of Effective Qubit Hamiltonian}
We begin from the microscopic Hamiltonian $\cH_{\rm el}$ defined in Eq.\ (6) of the main text which we recast as 
\begin{equation}\label{hh1}
H_\bk=\begin{pmatrix} h_d & T \\
T^\dagger & h_s
 \end{pmatrix},
\end{equation}
where $h_d=\xi_\bk\tau^z+\Delta_\bk\tau^x$ and $T=t_\bk\tau^z$.
We wish to `integrate out' the gapped $s$ degrees of freedom and find the effective theory for the electrons in $d$SC. The simplest way to accomplish this goal is to perform a unitary transformation $H\to\tilde{H}=e^{-S}H e^S$ with $S$ chosen such that $\tilde{h}_d$ and  
$\tilde{h}_s$ in the transformed Hamiltonian are decoupled, {\em i.e.\ } $\tilde{T}=0$. In this basis the $s$ degrees of freedom can be integrated out trivially. Following the steps outlined in Ref.~\cite{Zhong2007}
we find, to second order in $T$, 
\begin{equation}\label{hhg2}
\tilde{h}_d=h_d-{1\over 2}\left[T h_s^{-1}T^\dagger+h_dT h_s^{-2}T^\dagger +h.c.\right] +{\cal O}(T^4).
\end{equation}
This result is equivalent to treating Hamiltonian (\ref{hh1}) perturbatively to order $T^2$.

We are primarily interested in the effect of $h_s$ in the vicinity of the Dirac nodes of $h_d$ where we expect the formation of a proximity gap. We may thus neglect the second term in the angular brackets -- it is proportional to $h_d$ and it therefore vanishes at the nodes. The leading correction is then easily evaluated and reads,   
\begin{equation}\label{hhg3}
\tilde{h}_d\simeq h_d+{t_\bk^2\over (\xi_\bk^s)^2+\Delta_s^2}\begin{pmatrix} -\xi_\bk^s & e^{i\varphi}\Delta_s \\
e^{-i\varphi}\Delta_s  & \xi_\bk^s
 \end{pmatrix}.
\end{equation}
We conclude that electrons near the Fermi surface will have their dispersion modified according to $\xi_\bk\to\tilde{\xi}_\bk\simeq \xi_\bk-t_\bk^2\xi_\bk^s/\Delta_s^2$. More importantly, coupling to the $s$ layer will also modify the gap function  
\begin{equation}\label{hhg4}
\Delta_\bk \to \tilde\Delta_\bk =\Delta_\bk +e^{i\varphi} m_s,
\end{equation}
with $m_s\simeq {t_\bk^2/\Delta_s}$. One thus obtains the Hamiltonian $\cH_{\rm eff}$ given in Eq.\ (7) of the main text.

For a fixed classical phase $\varphi$ the energy spectrum of $\cH_{\rm eff}$ reads $E_\bk=\sqrt{\tilde{\xi}_\bk^2+|\tilde\Delta_\bk|^2}$ where
\begin{equation}\label{hhg5}
|\tilde\Delta_\bk|^2 =\underset{D_\bk^2}{\underbrace{\Delta_d^2\cos^2{2\alpha_\bk}+m_s^2}} + \underset{x_\bk}{\underbrace{2\Delta_d m_s\cos{2\alpha_\bk}\cos{\varphi}}}. 
\end{equation}
We next derive the effective potential for the phase variable $U(\varphi)$ by expanding the condensation energy of the system
\begin{equation}\label{hhg6}
E_g=\sum_\bk\left[|\tilde{\xi}_\bk|-\sqrt{\tilde{\xi}_\bk^2+|\tilde\Delta_\bk|^2} \right] 
\end{equation}
in the powers of $x_\bk$. Assuming a simple parabolic dispersion $\tilde{\xi}_\bk=\hbar^2k^2/2m-\mu$ we find
\begin{align}
E_g=&N_F\int_0^{\epsilon_c} d\xi\int_0^{2\pi}d\alpha
\biggl[(\xi-\sqrt{\xi^2+D^2}) \label{hhg7} \\
&-{x\over 2\sqrt{\xi^2+D^2}} 
+{x^2\over 8(\xi^2+D^2)^{3/2}} +O(x^3)\biggr], \nonumber
\end{align}
where $\epsilon_c$ is the energy cutoff and $N_F=\hbar^2L^2/2\pi^2m$ is the normal-state density of states.

The leading contribution to $U(\varphi)$ comes from the $x^2$ term, and upon extending $\epsilon_c\to\infty$ and performing the $\xi$ integral becomes
\begin{equation}\label{hhg8}
U(\varphi)=\cos^2{\varphi}\ m_s^2 N_F \int_0^{2\pi}d\alpha
{\cos^2{2\alpha}\over \cos^2{2\alpha} + (m_s/\Delta_d)^2}.
\end{equation}
We observe that  $U(\varphi)$ is $\pi$-periodic as expected for the $d/s$ junction. Once again this is strictly true if both layers respect the $C_4$ symmetry. It is easy to see that if the symmetry were lowered to $C_2$ say, the $x$ term in the expansion \eqref{hhg7} would generally be non-vanishing, resulting in a $\cos{\varphi}$ contribution to the potential.  It is also worth noting that the $\cos^2{\varphi}$ comes with a positive coefficient, leading to two minima at $\varphi=\pm\pi/2$. 

We note that
nodal regions, {\em i.e}.\ $\alpha\to \pm \pi/4$, make vanishing contribution to the integral in Eq.\ \eqref{hhg8}. A valid strategy in obtaining the low-energy effective theory, therefore, will be to integrate out $d$ quasiparticles outside the nodal regions but retain in the theory nodal Dirac fermions. This leads to an effective model for the nodal quasiparticles coupled to the qubit described by the Hamiltonian  $H_{\rm eff} = H_\varphi + H_{\rm qp}$ with 
\begin{align}
H_{\varphi} & = 4E_C(\hat{n}-n_g)^2 + E_J\cos 2\varphi+\delta U(\varphi)
\\
H_{\rm qp} &= \int_\bk \psi_\bk^\dagger \left[(v\bk-m_s\hat{x}\cos\varphi)\cdot\btau +m_s\sin\varphi\tau^z\right]\psi_\bk^{\vphantom\dagger}
\label{eq:Heff}
\end{align}
Here $\delta U(\varphi)$ denotes various $2\pi$-periodic contributions that we discuss in the main text and generally regard as small compared to $E_J$. For the sake of simplicity we shall assume a single species of Dirac fermions with isotropic velocity $v$, as indicated in Eq.\ \eqref{eq:Heff}.

Note that $n_\bk=\psi_\bk^\dagger\psi_\bk^{\vphantom\dagger}$ is conserved for each $k$, and the physical Hilbert space is $n_\bk=1\ \forall \bk$, so we can reduce the quasiparticle modes to effective spins: ${\bm S}_\bk = \psi_\bk^\dagger\btau\psi_\bk^{\vphantom\dagger}$.

\section{Effect of Quasiparticles I: Perturbative approach}
We next investigate the role of Bogoliubov quasiparticle excitations. For fixed value of $\varphi$, these quasiparticles are gapped except for $\varphi=0,\pi$, where they become gapless at four nodes on the Fermi surface. How do quantum fluctuations in $\varphi$ effect these nodal quasiparticles? 

To address this question, we focus on a single node, and project $H_{\rm qp}$ to the qubit subspace, i.e., the subspace spanned by the two lowest energy levels of the phase Hamiltonian $H_\varphi$. This projection will allow us to examine the fate of the Dirac quasiparticles in the presence of the fluctuating phase $\varphi$ and understand the back-action of quasiparticle on the phase dynamics.      

To carry out the projection we will require the matrix elements of $\sin{\varphi}$ and $\cos{\varphi}$ between the eigenstates of $H_\varphi$.
It will be useful to introduce the characteristic scales governing harmonic fluctuations of $\varphi$ around the minima of $U(\varphi)$
which are obtained by a quadratic approximation of $U$ near the minima.
The characteristic frequency of these oscillations is
\begin{align} 
    \omega_0 = 4\sqrt{E_JE_C},
\end{align}
and the corresponding RMS fluctuation amplitude is
\begin{align}
    \sigma_\varphi = (E_C/E_J)^{1/4}.
\end{align}
We are interested in the transmon regime: $E_J\gg E_C$, where the wave-function for the phase, $\varphi$ is well-localized near the minima at $\varphi=\pm \pi/2$, and the tunnel-splitting $\Delta E \sim  (E_C^3E_J)^{1/4}e^{-\sqrt{2E_J/E_C}}$ between the ground-states is small compared to $\omega_0$.

Let us consider $\eta=0$ where $H_\varphi$ conserves the parity of Cooper pairs, $\cP=(-1)^{\hat{n}}$.
We also focus on the particle-hole symmetric point $n_g=0$, where $H_\varphi$ has a charge-conjugation symmetry: 
\begin{align}
    \cC:&~e^{i\varphi}\mapsto e^{-i\varphi}, \nonumber\\
    \cC:&~n\mapsto -n.
\end{align}
{Parity conservation implies that we can label eigenstates of $H_\varphi$ by eigenvalues $P=\pm 1$ of the parity operator $\cP$. Denote the ground state of $H_\varphi$ in the symmetric ($P=+1$) as $|+\>$, and the ground-state in the antisymmetric ($P=-1$) sector as $|-\>$, and their respective eigenvalues as $E_S$, $E_A$. }

When the splitting between the qubit states, $\Delta E=E_A-E_S$ is much less than $m_s$, we can approximately project $H_{\rm qp}$ into the low-energy subspace spanned by $|\pm\>$. 
For this purpose, we will be interested in matrix elements of $e^{\pm i\varphi}$ in this subspace.
Noting $\cP|P\> = P|P\>$, we see that $e^{\pm i\varphi}$ has only off-diagonal elements in the $P$ basis $\<P|e^{i\varphi}|P\> =P^2\<P|\cP e^{i\varphi}\cP|P\> = -\<P|e^{i\varphi}|P\>=0$.
Charge conjugation symmetry implies: $\<-P|e^{i\varphi}|P\> = \<-P|\cC^\dagger e^{-i\varphi} \cC|P\> = -\<-P|e^{-i\varphi}|P\>$, such that $\<-P|\cos\varphi|P\>=0$, and the only non-zero matrix element is $\<-|\sin\varphi|+\>$.

This matrix element can be evaluated numerically or estimated by a tight-binding type approximation detailed below.
For a particle that is exactly localized at the potential minima $\varphi=\pm \pi/2$, we would have: $|\pm\> = \frac{1}{\sqrt{2}}\(|\varphi=\pi/2\>\pm |\varphi=-\pi/2\>\)$, and $\<+|\sin\varphi|-\> = 1$.
Gaussian fluctuations of $\varphi$ around the minima reduce the matrix element by a Debye-Waller factor that we now estimate.
In the weak-tunneling limit ($\Delta E\ll \omega_0$), we can approximate the qubit wave-functions by an even or odd superposition of Gaussian wave packets centered near $\varphi\approx \pm \pi/2$, 
\begin{align}
    |\pm\>\approx \frac{1}{\sqrt{2}\sqrt{2\pi \sigma_\varphi^2}}\[e^{-(\varphi-\pi/2)^2/4\sigma_\varphi^2}\pm e^{-(\varphi+\pi/2)^2/4\sigma_\varphi^2}\],
\end{align}
to obtain $\<-|\sin\varphi|+\>\approx A_{\rm DW}$, where we defined the Debye-Waller factor
\begin{align}\label{DW}
    A_{\rm DW}&\approx e^{-\frac12\<\delta \varphi^2\>} = e^{-\frac12 \sigma_\varphi^2} = 
    \exp\[-\frac12\sqrt{\frac{E_C}{E_J}}\].
\end{align}
\begin{figure}[t]
\includegraphics[width = 8.6cm]{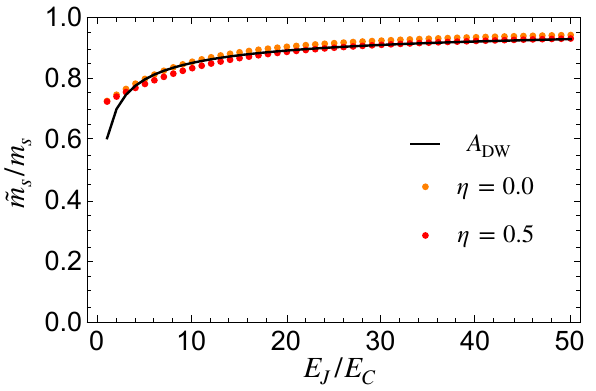}
\caption{The matrix element $\tilde{m}_s/m_s=\<-|\sin\varphi|+\>$ calculated numerically using eigenfunctions of $H_\varphi$ (dots) compared to the Debye-Waller factor Eq.\ \eqref{DW}. 
}\label{figS1}
\end{figure}
Projecting (\ref{eq:Heff}) into the low energy subspace of the qubit $\{|+\>,|-\>\}$, one then obtains:
\begin{align}\label{heff}
H_{\rm eff} \approx -\frac{\Delta E}{2}\sigma^z + \int_k\[v\bk\cdot {\bm S}_k + \tilde{m}_s \sigma^x S^z_k\]
\end{align}
where $\bsig$ are Pauli matrices for the qubit, e.g. $\sigma^z = |+\>\<+|-|-\>\<-|$. Also $\tilde{m}_s = A_{\rm DW} m_s$ is the  quasiparticle mass (renormalized by the Debye-Waller factor), and we have omitted additive constants.
In the transmon regime, $A_{\rm DW}\approx 1-\frac12\sqrt{\frac{E_C}{E_J}}$ is close to unity, and the renormalized quasiparticle mass, $\tilde{m}_s$ is only slightly smaller than the bare mass, $m_s$. Fig.\ \ref{figS1} shows $A_{\rm DW}$ compared to the result of a numerical calculation using the actual eigenfunctions of the phase Hamiltonian $H_\varphi$. A good agreement is seen for all $E_J/E_C\gtrsim 5$. We may conclude that in this regime the phase tunneling only results in a very modest suppression of the bare quasiparticle mass.

The model defined by Hamiltonian \eqref{heff} becomes solvable in the limit of vanishing qubit splitting ($\Delta E$ negligible compared to the quasiparticle energies). There are two degenerate sectors labeled by $\sigma^x=\pm 1$. The corresponding ground states in each sector are
\begin{align}
|G_\pm\> =   |\pm\>_\sigma\otimes\prod_\bk |v_\bk^\pm\>,
\end{align}
where $|\pm\>_\sigma$ are eigenstates of $\sigma^x$ while $|v_\bk^\pm\>$ denote eigenspinors of 
$h_\bk=\[v\bk\cdot {\bm S}_k + \tilde{m}_s \sigma^x S^z_k\]$
with a negative eigenvalue. In each sector, the quasiparticle excitations above the ground state are gapped, with energies $2\e_k$ with: $\e_k = \sqrt{(vk)^2+\tilde{m}_s^2}$, independent of $\sigma^x$. 
Crucially, the quasiparticle gap is non-vanishing for all $\bk$, and only slightly reduced from $m_s$ (by $A_{\rm DW}$) due to quantum zero-point motion of the phase.

To estimate the back-action of the quasiparticles on the qubit, we can consider perturbing $|G_\pm\>$ to lowest order in  $\Delta E \sigma^z$:
\begin{align}
\Delta \tilde{E} = \Delta E |\<G_+|\sigma^z|G_-\>| = \Delta E \prod_\bk |\<v_\bk^+|v_\bk^-\>|^{2_S}
\end{align}
where the factor of $2_S$ arises from the spin-degeneracy of the quasiparticles.
Evaluating the product over momenta using the continuum approximation $\prod_\bk g_\bk = \exp[\frac{L^2}{(2\pi)^2}\int d^2k \ln{g_\bk}]$ with 
\begin{align}\label{gk}
g_k=|\<v_\bk^+|v_\bk^-\>|^{2}={(vk)^2\over (vk)^2+\tilde{m}^2_s}
\end{align}
gives~\footnote{Note that in a finite system, the minimum $k$ is bounded below by $\sim 1/L$, so $k=0$ does not appear in the product, so that the product is non-vanishing for any finite $L$ even though $\lim_{k\rightarrow 0}< v_k^+|v_k^->=0$.}:
\begin{align}\label{de1}
\Delta\tilde{E} \approx  \Delta E\  \exp\left[-\frac{L^2\tilde{m}_s^2}{2\pi v^2}f(\tilde\xi \Lambda)\right]
\end{align}
where $f(x) = \frac12\ln (1+x^2) -x^2\ln \frac{x}{\sqrt{x^2+1}}$, $\tilde{\xi} = {v/\tilde{m}_s}$ is the renormalized superconducting coherence length, $L$ is the linear dimension of the system, and $\Lambda$ is a UV cutoff on the order of the lattice spacing.
Note that, the bare qubit splitting also scales exponentially in system volume, as $\Delta E \sim e^{-(L/\xi_J)^2}$ where 
$\xi_J = \(e^2/4E_JE_C\)^{1/4}$  is the Josephson coherence length, so the Dirac quasiparticles simply renormalize the coefficient of the $L^2$ dependence:
\begin{align}
\frac{1}{\xi_J^2}\rightarrow \frac{1}{\xi_J^2} + \frac{f(\tilde{\xi}\Lambda)}{2\pi \tilde{\xi}^2}
\end{align}

We conclude that (i) in the transmon regime the phase dynamics does not alter the fact that quasiparticles are gapped due to the $d/s$ proximity effect (although the gap is slightly reduced from its mean-field value), and (ii) the back-action of the low-energy quasiparticles on the qubit is limited to a weak renormalization of the qubit energy splitting as quantified by Eq.\ \eqref{de1}.

\section{Effect of quasiparticles II: Instanton Gas Approximation}
In this appendix, we provide a complementary analysis of the effect of nodal quasiparticles on the $d$-mon qubit using a semiclassical approach in which tunneling between qubit states is modeled as instanton events in an imaginary time path integral. Again, we focus on the case where there is no external flux ($\eta=0$). 
The object of interest will be the imaginary time propagators:
\begin{align}
    G_\pm &= \<\pm \varphi_0|e^{-\beta H}|\varphi_0\>
\end{align}
where $\varphi_0 = \frac{\pi}{2}$ denotes the minima of the $E_J\cos2\varphi$ potential, and $|\pm \varphi_0\>$ denote the mean-field state with $\varphi=\varphi_0$, and the corresponding ground-state of the quasiparticles of $H_{\rm qp}$ with fixed $\varphi=\pm \varphi_0$.

We can write $G_\pm$ in terms of a coherent state path integral $G_\pm = \int D\varphi D[\bar\psi,\psi] e^{-S[\varphi,\bar\psi,\psi]}$ with $S= S_\varphi + S_{\rm qp}$ and 
\begin{align}
    S_\varphi &= \int_0^\beta d\tau \[\frac{1}{4E_C}(\d_\tau \varphi)^2+E_J(\cos{2\varphi} +1)
    \]\\ 
    \nonumber
    S_{\rm qp} &= \int_0^\beta d\tau \int_\bk \bar\psi_\bk \[\d_\tau + (v\bk\cdot\btau +m_s\sin\varphi\tau^z)\]\psi_\bk.
\end{align}
Here, as in the previous section, we have neglected the $m_s\hat{x}\cos\varphi$ term, which averages to zero in the low energy subspace of the qubits.

The standard semiclassical ``dilute instanton gas" approximation involves focusing on small fluctuations of $\varphi = \varphi_c+\delta\varphi$ around instanton-type solutions to its classical equations of motion 
\begin{align}\label{eqm}
\d_\tau^2\varphi_c = U'(\varphi_c), \ \ U(\varphi) = E_J(\cos{2\varphi}+1). 
\end{align}
This imaginary-time equation of motion resembles a real-time equations of motion for a particle in an inverted potential: $U(\varphi) = -E_J(\cos{2\varphi}+1)$, and classical solutions include instanton trajectories where $\varphi$ rolls between maxima of $U(\varphi)$ at $\varphi=\pm\pi/2$. 
Since the potential is periodic, there are paths in which $\varphi$ can roll from $\frac{\pi}{2} \rightarrow \frac{\pi}{2}\pm \pi$.

Denote as $S_0$, the classical action cost, $S_\varphi[\varphi_c]$ of an instanton solution to Eq.\ \eqref{eqm}:
\begin{align}
    S_0 = 
    \int_{-\pi/2}^{\pi/2} d\varphi \sqrt{\frac{U(\varphi)}{4E_C}}
   = \sqrt{\frac{2E_J}{E_C}}.
\end{align}
(note this expression also follows from the WKB approximation to the tunneling amplitude).
\begin{widetext}
The weight of a single instanton configuration is $Ke^{-S_0}$, where~\cite{AltlandSimons}: 
$K = c\omega_0 \sqrt{\frac{S_0}{2\pi}}$
represents Gaussian fluctuations around the classical solution (for a quartic potential, the constant is $c=\frac{1}{12}$~\cite{ZinnJustin}).
The typical time between such instanton events is then $\Delta T_I \sim \Delta E^{-1}\sim e^{\sqrt{2E_J/E_C}}$.
In the transmon regime, $E_J\gg E_C$, this time-scale is much larger than the characteristic time scale of the instanton event itself, which is of order $T_I\sim 1/\omega_0\sim \sqrt{E_C/E_J}$.
In this limit, we may then approximate the path integral via a dilute ``gas" of some number, $N$ of instantons, with negligible amplitude for overlapping instantons:
\begin{align}\label{igas}
    G_{\pm} \approx  \sum_{N_++N_-={\rm even/odd}}\frac{1}{N_+!N_-!}\int d\tau^+_1\dots d\tau^+_{N_+} d\tau^-_1\dots d\tau^-_{N_-} (Ke^{-S_0})^N \int D[\bar\psi,\psi] e^{-S_{\rm qp}[\varphi_c]}
\end{align}
where $\varphi_c$ is the classical solution for $N_{\pm}$ instantons where $\varphi$ changes from $\varphi\rightarrow \varphi\pm \pi$, occuring at the times $\tau_i^{\pm}$. 
\end{widetext}
This expression differs from that for a double-well potential, due to the presence of two paths between the minima: $\varphi = \pi/2 \rightarrow \pi/2 \pm \pi$ in the periodic $\sim\cos2\varphi$ potential.

In the absence of the quasiparticle modes, the partition function would become: $G_+ = \cosh \beta 2Ke^{-S_0}$, and $G_- = \sinh \beta 2Ke^{-S_0}$, where the factor of $2$ difference from the usual double-well potential arises since there are two paths for $\varphi$ to tunnel between $\pm \frac{\pi}{2}$ (increasing or decreasing by $\pi$) in the periodic potential. To interpret these expressions, consider large $\beta$ where $e^{-\beta H}$ approximately projects into the qubit subspace spanned by states even and odd Cooper-pair parity ground-states: $|\pm\>$ (as described in the previous section), i.e. 
\begin{align}
    G_\pm &\approx \<\pm \varphi_0|e^{-\beta H}\sum_{P\in \{S,A\}}|P\>\<P|\varphi_0 
    \nonumber\\
    &\approx \<\pm \varphi_0|+\>\<+|\varphi_0\>e^{-\beta E_S} + 
    \<\pm \varphi_0|-\>\<+|\varphi_0\>e^{-\beta E_A}
    \nonumber\\
    &= e^{-\beta (E_S+E_A)/2}
    \begin{cases} 
    \cosh \beta \Delta E &; +~{\rm case} \\
    \sinh \beta \Delta E &; -~{\rm case}
    \end{cases},\label{Gpm}
\end{align}
where $\Delta E = (E_A-E_S)/2$ is the qubit splitting. Comparing these expressions, we identify the bare qubit splitting:
\begin{align}
    \Delta E = Ke^{-S_0}\sim (E_J^3 E_C)^{1/4}e^{-\sqrt{2E_J/E_C}}
\end{align}
which is consistent with Eq.\ (5)  derived by exact methods.

We now turn to the effect of the quasiparticles near the nodes of the $d$-wave gap, which have so far been neglected.
We can divide the quasiparticle modes for different $k$ into two limiting cases by comparing the instanton time $T_I$ to the transition rate between quasiparticle states: those for which the instantons are approximately adiabatic, and those for which the instantons are sudden.
We will treat these limiting cases separately.
Intermediate cases that are neither adiabatic nor sudden are more complicated to treat, and we instead assume that the results smoothly interpolate between the extremes.

\subsection{Adiabatic modes}
Take the imaginary time coordinate, $\tau \in [-\beta/2,\beta/2]$, and consider a single instanton event, $\varphi_1(\tau)$ that we take for convenience to occur at $\tau_1=0$.
The problem of a single quasiparticle mode in this instanton background takes the form the classic Landau-Zener (LZ) problem, but in imaginary time.
Denoting the instantaneous quasiparticle energies for a fixed $\varphi$ as $E(\varphi) = \sqrt{(vk)^2+m^2_s\sin^2\varphi}$, the LZ diabatic parameter is:
\begin{align}
    g = \frac{||\d_\tau h_{\rm qp}||}{\min_\varphi |E(\varphi)|^2}\approx \frac{m_s}{(vk)^2T_I}
\end{align}
Quasiparticle modes with $g\ll 1$, see the instanton dynamics as slow, and nearly-adiabatically follow the instantaneous ground-state as a function of $\varphi_c$.
The adiabatically changing quasiparticle energy renormalizes the instanton action cost as: $S_0\rightarrow S_0 +\Delta S_A$, with:
\begin{align}
    \Delta S_A &\approx {\sum_k}' \int \[E_k(\pi/2)-E_k(\varphi_c(\tau)\]d\tau
    \nonumber\\
    &\lesssim 
    {\sum_k}' T_I\[\sqrt{(vk)^2+m^2_s}-|vk|\]
    \nonumber\\
    &\approx 
    \frac{L^2T_I}{2\pi} \int_{k_g}^{\Lambda} kdk \(\sqrt{(vk)^2+m^2_s}-vk\)
    \nonumber\\
    &\approx 
    \frac{L^2m_sT_I\Lambda}{4\pi v^2}
\end{align}
where $\sum_k'$ indicates a restricted sum over quasiparticle modes that satisfy the adiabatic condition $g\ll 1$, $\Lambda$ is a UV cutoff, and $k_g$ is defined as the $k$ for which the adiabatic approximation fails: $k_g\approx \sqrt{m_s/T_Iv^2}$.

We note, in passing, that corrections to the adiabatic approximation can be systematically obtained via time-dependent perturbation theory by moving to the instaneous quasiparticle eigenstate basis for fixed $\varphi$, defined by rotation $U(\varphi) = e^{-\frac{i}{2}\tan^{-1}(m_s\sin\varphi/vk)\tau^y}$. In this frame, the quasiparticle (single-particle) Hamiltonian becomes $\tilde{h}_{\rm qp} = E(\theta)\tau^z + \mathcal{A}\d_{\tau}\varphi$, where we have introduced the non-Abelian Berry connection: $\mathcal{A} = -iU^\dagger \d_\varphi U = \frac12 \frac{m_svk}{E(\varphi)^2}\cos\varphi \tau^y$.
However, given that we do not expect these effects to be qualitatively important, we do not pursue this approach further.

From these considerations, we see that the adiabatic modes simply renormalize the potential for the phase. Let us absorb this renormalization into the definition of the qubit parameters: $E_{J,C}$, and focus on the remaining quasiparticle modes that violate the adiabatic condition.

\subsection{Sudden modes}
At the other extreme are low-energy quasiparticle modes for which the instanton is sudden, and the quasiparticle wave-function does not have time to appreciably change during the instanton event.
For these modes, we can approximate each instanton profile by a sudden step function: $\varphi(\tau) = \sum_{i} \frac{\pi}{2}{\rm sgn}(\tau-\tau_i)$, so that the effective time evolution for $Z_{\rm qp}[\varphi] = \int D[\bar\psi,\psi] e^{-S_{\rm qp}(\psi,\varphi)}$ takes the form of piecewise constant time evolution with $\varphi = \pm \frac{\pi}{2}$.

To evaluate the instanton action in the sudden limit, note the following observations.
For large, $\tau$, $e^{-\tau h}\approx e^{-\tau E_0}|E_0\>\<E_0|$ is approximately proportional to a projector onto the quasiparticle ground state, $|E_0\>$ of the single-particle Hamiltonian $h$. 
Hence the product of two piecewise constant imaginary-time propagators: $e^{-\tau_2 h_2}e^{-\tau_1h_1}\approx \[e^{-(\tau_2E_0^2+\tau_1E_0^1)}\<E_0^1|E_0^2\>\] |E_0^2\>\<E_0^1|$, is approximately proportional to the overlap of wave-functions.

Now apply these approximations  to the instanton gas expansion (Eq.~\ref{igas}). Note that, in the transmon regime, the time between instantons $\sim e^{S_0}$ is much larger than the quasiparticle gap $\gtrsim m_s$, so that we can use the long-time limit formulas above.
Note also that the quasiparticle ground-state energy $E(\varphi)$ is the same for $\varphi = \pm \frac{\pi}{2}$. The result is:
\begin{align}
    G_\pm|_{\rm sudden} &\approx \sum_{N={\rm even/odd}}\frac{1}{N!}\(2\beta e^{-S_0}\)^N 
    |\<v_+|v_-\>|^N
    \nonumber\\
    &\approx 
 e^{-\beta\int_{k}E_k}
 \begin{cases}
        \cosh \[\beta 2e^{-S_0}\prod_\bk' g_k\] 
        \\
        \sinh \[\beta 2e^{-S_0}\prod_\bk' g_k\]
    \end{cases}
\end{align}
and the prime indicates restriction in the product to sudden modes $|\bk|<k_g$.
Comparing to expression \eqref{Gpm} above, we recover the result from the previous section that the qubit tunneling is renormalized by a product of wave-function overlaps:
\begin{align}
    \Delta \tilde{E} = \Delta E \prod_{k< k_g} |\<v_\bk^+|v_\bk^-\>|^2.
\end{align}
Here, we see that the cutoff for the momentum sum should be placed at the characteristic $k$-scale at which the sudden instanton approximation breaks down, rather than the lattice scale. Note that, as computed below Eq.\ \eqref{de1} above, the resulting renormalization is only logarithmically sensitive to precise value of the cutoff.

To summarize, the results of the more complicated instanton-gas calculation gratifyingly reproduce the conclusions  of the simpler effective Hamiltonian approach of the previous section, with some minor addenda such as additional renormalization of the instanton action due to the adiabatic modes. Both approaches consistently conclude that the nodal quasiparticles are fully gapped, even during tunneling events where the qubit phase passes (in transient/virtual quantum motion) through $\varphi=0,\pi$.

\subsection{Instanton gas approximation of $\<-|\sin\varphi|+\>$}
As a final application of the semiclassical path integral methods, we revisit the tight-binding approximation for $\<-|\sin\varphi|+\>$ with instanton gas methods.
Since we have seen that quasiparticles merely renormalize the instanton action, we ignore them in this analysis.
Denote the potential minima of $H_\varphi$ as $\varphi_0=\pm \pi/2$, then we may approximate the low-energy states $|\pm\>$ by:
\begin{align}
|\pm\> = \lim_{\beta\rightarrow \infty} \frac{1}{\mathcal{N}^{\pm}_{\beta}}\(|\varphi_0\>\pm |-\varphi_0\>\).
\label{eq:SA}
\end{align}
This formula follows from the observations that $e^{-\beta H}$ acts as a projector onto low energies for large $\beta$, and that $ |\varphi_0\>\pm |-\varphi_0\>$ have Cooper pair parities $\pm 1$ respectively so that by symmetry they only have overlap with $|\pm\>$ respectively.
Here, $\mathcal{N}_\beta^{\pm} = |\<S/A|\pm\>|$ are normalization constants that will be determined shortly.

\begin{widetext}
Inserting the expression for (\ref{eq:SA}), and exploiting the parity symmetry of the resulting matrix elements, gives:
\begin{align}
\<-|e^{i\varphi}|+\> = \lim_{\beta\rightarrow \infty} \frac{2}{\mathcal{N}_\beta^+\mathcal{N}_\beta^-}\[\<\varphi_0|e^{-\beta H}e^{i\varphi}e^{-\beta H}|\varphi_0\> - \<-\varphi_0|e^{-\beta H}e^{i\varphi}e^{-\beta H}|\varphi_0\>\].
\end{align}
\end{widetext}

We can cast these expressions as a path integral:
\begin{align}
\<\pm \varphi_0|e^{-\beta H}e^{i\varphi}e^{-\beta H}|\varphi_0\>  = \int D\varphi|_{\varphi(-\beta)=\varphi_0}^{\varphi(\beta)=\pm \varphi_0} e^{-S[\varphi]+i\varphi(\tau=0)}, 
\end{align}
which can then be evaluated semiclassically by expanding $\varphi \approx \varphi_{\rm cl}+\delta \varphi$ where $\varphi_{\rm cl}$ represent a solution to the classical equations of motion in imaginary time with $N$ instantons ($N$ and the locations of the instantons are then summed over), and $\delta \varphi$ represents quadratic fluctuations about this classical solution.

The insertion of the phase operator, $e^{i\varphi(\tau=0)}$, causes all of the configurations with a non-zero number of instantons to vanish. To see this, note that, in the non-interacting instanton gas approximation, $e^{i\varphi(\tau=0)}$ is freely oscillating between $\pm i$ depending on whether there are an even or odd number of instantons prior to $\tau=0$.

Thus the numerator reduces to the contribution from the zero instanton sector, and simply gives $2iA_{\rm DW}$ as obtained in the simple tight-binding approximation described above.

To conclude, we need to also evaluate the normalization factors:
\begin{align}
(\mathcal{N}_\beta^{\pm})^2 &= 2\(\<\varphi_0|e^{-2\beta H}|\varphi_0\>\pm 2{\rm Re}\<-\varphi_0|e^{-2\beta H}|\varphi_0\>\) 
\nonumber\\&
\approx 2\(\cosh 2\beta e^{-S_0} \pm \sinh 2\beta e^{-S_0}\)
\end{align}
where the second line follows from the standard instanton gas approximation, and $S_0$ is the action cost for a single instanton configuration (including Gaussian fluctuations around the classical solution).
Then, we have, simply $\mathcal{N}_\beta^{+}\mathcal{N}_\beta^{-} = 2 \(\cosh^2 (2\beta e^{-S_0})-\sinh^2(2\beta e^{-S_0})\)^{1/2} = 2$.
Assembling these intermediate results reproduces the result $\<-|\sin \varphi|+\> \approx A_{\rm DW}$ obtained by simpler methods above.

\end{document}